\documentclass[showpacs, amsmath, amssymb,twocolumn]{revtex4}

\usepackage{color}
\usepackage{amsfonts}
\usepackage{amsbsy}
\usepackage{latexsym}
\usepackage{mathrsfs}
\usepackage{bbm}
\usepackage{bm}
\usepackage[dvipdfm]{graphicx}

\newcommand{\be}{\begin{eqnarray}}
\newcommand{\ee}{\end{eqnarray}}
\def\({\left(}
\def\){\right)}
\def\[{\left[}
\def\]{\right]}

\def\C{\mathbb{ C}}

\newcommand{\bra}[1]{\left\langle #1 \right|}
\newcommand{\ket}[1]{\left| #1 \right\rangle}
\newcommand{\set}[2]{\left\{ #1\big\vert #2 \right\}}

\newcommand{\Tr}{\mathrm{Tr}}
\newcommand{\sla}[1]{\rlap{\kern .15em /}#1}
\newcommand{\s}{\hspace{2mm}}

\newcommand{\ot}{\otimes}

\newcommand{\op}{\oplus}

\begin{document}
\title{Entanglement of Indistinguishable Particles}
\author{
Toshihiko Sasaki$^{1}$, Tsubasa Ichikawa$^2$, and Izumi Tsutsui$^{1, 3}$
}

\affiliation{
$^1$Department of Physics,  
University of Tokyo, 7-3-1 Hongo, Bunkyo-ku, Tokyo 113-0033, Japan\\
$^2$Research Center for Quantum Computing, Interdisciplinary Graduate School of Science and Engineering, 
Kinki University, 3-4-1 Kowakae, Higashi-Osaka, Osaka 577-8502, Japan\\
$^3$Theory Center, 
Institute of Particle and Nuclear Studies, High Energy Accelerator Research Organization (KEK), 1-1 Oho, Tsukuba, Ibaraki 305-0801, Japan
}

\begin{abstract}
We present a general criterion for entanglement of $N$ indistinguishable particles decomposed into arbitrary $s$ subsystems based on the unambiguous measurability of correlation.   Our argument provides a unified viewpoint on the entanglement of indistinguishable particles, which is still unsettled despite various proposals made mainly for the $s=2$ case.    Even though entanglement is defined only with reference to the measurement setup, we find that the so-called  i.i.d.~states form a special class of bosonic states which are universally separable.  

\end{abstract}

\maketitle

\section{Introduction}

Since its first recognition in the seminal EPR and Schr{\" o}dinger's papers \cite{EPR, Sch}, quantum entanglement has long been seen as 
the most distinctive trait of quantum theory.  Notably, it underlies nonlocal correlation in composite physical systems, invoking various conceptual questions on the foundation of physics and,  
at the same time, offers
a key resource for quantum information sciences.   In view of this, we find it rather puzzling that the very notion of entanglement still eludes a formal, let alone intuitive, understanding, especially when the system admits no apparent decomposition into subsystems.   This occurs typically in systems of indistinguishable particles ({\it i.e.}, fermions or bosons) with which actual realizations of entanglement  -- via photons, electrons or composite particles such as hydrogen atoms -- have been implemented mostly today.

To see the nontrivial nature of entanglement, take, for example, the familiar $N=2$ particle Bell states, 
\begin{equation}
\ket{\Psi} = {1\over \sqrt2} (\ket{0}_1\ket{1}_2 \pm \ket{1}_1\ket{0}_2),
 \label{bellst}
\end{equation}
with $\ket{0}_k$ and $\ket{1}_k$ being the orthonormal qubit states of particle $k = 1, 2$.  These are prototypical entangled states for distinguishable particles, but if the particles are indistinguishable, the labels $k$ are no longer usable for classifying the measurement outcomes to define the correlation.  A remedy for this is to consider remotely separated particles by introducing the position degrees of freedoms directly for (\ref{bellst}), but this does not yield any nontrivial correlation, a property known as the cluster separability \cite{WC63, Peres93}.
Clearly, a physically motivated and mathematically solid definition of entanglement is needed for general composite systems including those of indistinguishable particles. 

Recently, Ghirardi {\it et al}.~\cite{GMW02, {GM03}} gave a possible definition of separability (non-entanglement) for $N$ particle systems based on the criterion that, if one can deduce a complete set of physical properties (CSP) pertaining to a subsystem, then the state is separable with respect to the subsystem and the rest.  
This criterion derives from the demand that, in a separable state, all physical quantities in the subsystem have elements of reality  in the EPR sense \cite{EPR}.     
Independently, Zanardi {\it et al}.~\cite{Zanardi01} presented a criterion for uncovering a tensor product structure (TPS) in the Hilbert space upon which entanglement can be defined.    The criterion demands the existence of subalgebras representing the observables of the subsystems, which are measurable, independent,  and complete to form the entire set of measurable observables in the system.    Yet another criterion has been proposed by Schliemann  {\it et al}.~\cite{SCKLL01} and others \cite{PY01} particularly for indistinguishable particles using the (Schmidt or Slater) rank of state decomposition, which is related to the standard measures of entanglement such as  the von Neumann entropy.

These criteria for (non)entanglement are rather different from each other and, 
not surprisingly, do not completely agree on deciding which states are separable, with an example being the $N = 2$ bosonic \lq independently and identically distributed\rq\ (i.i.d.) state $\ket{\phi}_1\ket{\phi}_2$ (for an attempt of reconciliation, see \cite{{GM04}}).
More recently, the present authors furnished a criterion for the decomposition of an $N$ fermionic system into $s$ arbitrary subsystems \cite{IST10},  where we find that the orthogonal structure introduced to distinguish the subsystems in \cite{GMW02} corresponds precisely to the choice of observables with which correlation is defined.  In other words, entanglement can be defined only {\it relative} to the measurement setup and it is highly non-unique \cite{Zanardi01}.   Under these circumstances, one is naturally led to ask if there is any coherent picture of entanglement prevalent among these criteria.  

The purpose of this article is to provide a positive answer to this.   Namely, we show that all these criteria can be  put into a larger perspective consisting of two descriptions of the system, one for the measurement outcomes and the other for the provisional states of the system.   
The gap between the two descriptions, which lies at the root of the apparent disagreement,  can be filled by an isomorphism between the two, providing a unified viewpoint of entanglement for indistinguishable particles.   
Unlike the previous analyses,  entanglement can be treated equally for the fermionic and bosonic cases here.  We also find that the i.i.d.~states 
for general $N$ form a special class of bosonic states which are universally separable irrespective of the choice of measurement setup.

\section{Entanglement in measurement-based description}

To define entanglement as an attribute to generate nontrivial correlation among subsystems, we first need an appropriate set of physical observables associated with the subsystems for which the correlation in their measurement outcomes can be evaluated unambiguously.  
To discuss the situation explicitly, we consider the case where the total system breaks into $s$ subsystems $\Gamma_1, \ldots, \Gamma_s$ and
assume that to each $\Gamma_i$ we have a complete set of commuting observables ${\cal C}_i$ which are all implementable in the measurement to  determine the state of the subsystem.   Let ${\cal L}_i$ be the set of observables (self-adjoint operators) containing the set ${\cal C}_i$.  The collection of
states of the subsystem $\Gamma_i$ describing the measurement outcomes form a Hilbert space $\mathcal{H}^{\rm mes}(\Gamma_i)$ in which ${\cal L}_i$ is represented irreducibly.  
Assuming further that 
the measurements of the observables ${\cal L}_i$ can be performed independently for all $i = 1, \dots, s$, we find that the  
set ${\cal L}$ of observables in the total system is given by ${\cal L} = \otimes_i {\cal L}_i$.   
Accordingly, the state space of the system describing the measurement outcomes is given by the tensor product,
\begin{equation}
\mathcal{H}^{\rm mes} = \bigotimes_{i=1}^s \mathcal{H}^{\rm mes}(\Gamma_i).
\label{ortps}
\end{equation}
The TPS of the total space $\mathcal{H}^{\rm mes}$ in (\ref{ortps}) allows us to define the entanglement by the
conventional way, that is, if the measured state  $\ket{\Psi} \in \mathcal{H}^{\rm mes}$ admits the product form 
\begin{equation}
\ket{\Psi} = \bigotimes_{i=1}^s \ket{\psi_i}_{\Gamma_i}, \quad 
\ket{\psi_i}_{\Gamma_i} \in \mathcal{H}^{\rm mes}(\Gamma_i),
 \label{orseps}
\end{equation}
 then it is separable; if not, it is entangled.   
Evidently, since the separable state (\ref{orseps}) yields definite outcomes for
the measurement of observables in a properly chosen set ${\cal C}_i$ in $\ket{\psi_i}_{\Gamma_i}$ for all $i$, it possesses a CSP  \cite{GMW02}.  

Meanwhile, in the total space $\mathcal{H}^{\rm mes}$ the observable $O_i \in {\cal L}_i$ is expressed by
\begin{equation}
     \widehat{O}_i = \bigotimes_{j=1}^{i-1}\mathbbm{1}_j\otimes O_i \otimes \bigotimes_{j=i+1}^s\mathbbm{1}_j 
 \label{ortildeO}
\end{equation}
where $\mathbbm{1}_j$ is the identity operator in $\mathcal{H}^{\rm mes}(\Gamma_j)$.  The aforementioned independence is then assured trivially by
\be
[    \widehat{O}_i,     \widehat{O}_j]=0
\qquad
{\rm for}
\qquad
i\neq j.
\label{oindcon}
\ee
The observable $\widehat O \in {\cal L}$ corresponding to the simultaneous measurement for the subsystems is then given by 
\be
\widehat O = \prod_{i=1}^s    \widehat{O}_i= \bigotimes_{i=1}^{s} O_i.
\label{orprodone}
\ee
Denoting the set of such operators by ${\cal T}^{\rm mes} \subset {\cal L}$, we see that  any $\widehat O \in {\cal T}^{\rm mes}$ has a factorized expectation value
for the separable state $\ket{\Psi} $ in (\ref{orseps}):
\begin{eqnarray}
\bra{\Psi}  \widehat{O} \ket{\Psi} = \prod_{i=1}^s \bra{\Psi}    \widehat{O}_i\ket{\Psi}.
 \label{facto}
\end{eqnarray}
The properties (\ref{oindcon}) and (\ref{orprodone}), together with the implementability assumption, constitute the formal conditions  
to realize a TPS in \cite{Zanardi01}.   Note that, in our measurement-based description, the TPS appears as a direct consequence of the construction.

\section{Entanglement in provision-based description}

The entanglement in the measurement-based description is related with the measurement outcomes directly, but the conventional treatment 
of indistinguishable particles employ the framework of the provisional Hilbert space of distinguishable particles for the description of states with appropriate restriction 
required by the statistics of the particles.   Here,  the description is not directly related to the measurement outcomes, and the restricted space of states does not admit a TPS in any obvious manner.   In physical terms, the measurement outcomes of observables, such as spin, cannot be attributed to those of a particular particle due to the indistinguishability, and the formal structure of the state fails to signify the correlation as exemplified by (\ref{bellst}).  To fill the gap, we need a prescription to bridge the two descriptions.

{}For definiteness, let us label the $N$ particles by the integer set $\mathcal{N}=\{1,2,\cdots, N\}$.   Each of the particles is characterized by an $n$-level state, {\it i.e.}, the state space of the $k$-th particle is $\mathcal{H}_k\cong\C^n$ for all $k$.  
Let $\{\ket{e_i}\}$ be a complete orthonormal basis in $\C^n$.
By the isomorphism among the constituent spaces ${\cal H}_k$, any pure state $\ket{\Psi}$ in the provisional space 
$\mathcal{H}= \bigotimes_{k\in \mathcal{N}} \mathcal{H}_k$ of the total system can be written as
\begin{equation}
 \ket{\Psi} = \sum_{i_1,i_2,\cdots,i_N}\Psi_{i_1i_2\cdots i_N} \bigotimes_{k=1}^N \ket{e_{i_k}}_k,
 \end{equation}
where $\Psi_{i_1i_2\cdots i_N}\in \mathbb{C}$ and $\{\ket{e_{i_k}}_k\}$ is the complete orthonormal basis in ${\cal H}_k$ isomorphic to $\{\ket{e_i}\}$.

To incorporate the indistinguishability of the particles, consider an element $\sigma\in\mathfrak{S}_N$ of 
the symmetric group $\mathfrak{S}_{N}$ associated with the permutation
$k \to \sigma(k)$.   In $\mathcal{H}$, the element
is represented  by a self-adjoint operator $\pi_\sigma$ with
\begin{equation}
 \pi_{\sigma} \ket{\Psi} = \sum_{i_1,i_2,\cdots,i_N}\Psi_{i_1i_2\cdots i_N} \bigotimes_{k=1}^N \ket{e_{i_k}}_{\sigma(k)}.
\end{equation}
From $\pi_\sigma$, both the symmetrizer and the antisymmetrizer are defined as
\begin{equation}
 \mathcal{S}=\frac{1}{N!} \sum_{\sigma\in \mathfrak{S}_{N}} \pi_{\sigma},
 \qquad
 \mathcal{A}=\frac{1}{N!} \sum_{\sigma\in \mathfrak{S}_{N}} \text{sgn}(\sigma) \pi_\sigma,
  \label{symop}
\end{equation}
where $\text{sgn}(\sigma)$ is the signature of the permutation $\sigma$.  
The Hilbert space of the total system of 
$N$ bosons (fermions) is the subspace of $\mathcal{H}$ consisting of symmetric (antisymmetric) states.  
Putting $\mathcal{X} = \mathcal{S}$ for bosons and $\mathcal{X} = \mathcal{A}$ for fermions, they are
\begin{equation}
\mathcal{H}_{\mathcal{X}} =  \left[ \mathcal{H}\right]_{\mathcal{X}} :=\set{\mathcal{X}\ket{\Psi}}{\ket{\Psi}\in \mathcal{H}}.
 \label{xsusp}
\end{equation}

To introduce the decomposition into subsystems in the total system, we consider
a partition $\Gamma$ of the integer set $\mathcal{N}$ into non-empty and mutually exclusive sets $\Gamma_i\subseteq{\cal N}$, 
\begin{eqnarray}
\Gamma = \left\{\Gamma_i\right\}_{i=1}^s,\s
\bigcup_{i=1}^s \Gamma_i=\mathcal{N},\s
\Gamma_i\cap \Gamma_j=\emptyset  \s
\,\, \text{for $i\neq j$}.
\end{eqnarray}
For specifying the subsystems of indistinguishable particles, only the cardinality $\vert \Gamma_i \vert$ of $ \Gamma_i$ matters.   
Note that there is no apparent TPS in $\mathcal{H}_{\mathcal{X}}$
with respect to $\Gamma$, and we need to somehow find an embedding of the measurement-based description in the provision-based description.  

This embedding is handled usually by considering the positions of particles to gain a fictitious distinction between the 
particles.   For the distinction to be unambiguous, the measurements of the subsystems should be performed remotely from each other, and this amounts to introducing an 
orthogonal decomposition in the 1-particle Hilbert space (after accommodating the position degrees of freedoms).   
More generally, the embedding requires an orthogonal structure $V$ which is a set of subspaces $V_i \subset \C^n$ mutually orthogonal to each other with respect to the innerproduct of $\C^n$,
\be
V=\{V_i\}_{i=1}^s, 
\qquad
V_i\perp V_j\s\,\, \text{for $i\neq j$}.
\label{ortst}
\ee
Together with the orthogonal complement,
\be
V_0=\(V_1\op V_2\op\cdots\op V_s\)^\perp,
\label{addone}
\ee
the set $V$ furnishes an orthogonal decomposition of $\C^n$.   The physical idea behind this is that these orthogonal spaces correspond to 
mutually independent measurement of subsystems such that, given a measurement setup, the subsystem $\Gamma_i$ is susceptible only for the measurement of particles $k \in \Gamma_i$ residing in $V_i$. 
If we denote the subspace $V_i$ in $\mathcal{H}_k$ by $V_i(\mathcal{H}_k) \subset \mathcal{H}_k$, then the actual Hilbert space 
describing the measurement outcomes for $\Gamma_i$ is given by
\begin{equation}
\mathcal{H}_{\mathcal{X}}(\Gamma_i, V_i) = 
\left[\bigotimes_{k\in \Gamma_i}V_i(\mathcal{H}_k) \right]_{\mathcal{X}}.
\label{hxsp}
\end{equation}
Here, as in (\ref{xsusp}),  the index $\mathcal{X}$ attached to $[*]$ implies that it is the subspace of $*$ invariant under $\mathcal{X}$ associated with the symmetry group $\mathfrak{S}_{K}$ with $K$ being the cardinality of $*$, and in  (\ref{hxsp}) we have $K = |\Gamma_i|$.
Clearly, 
$\mathcal{H}_{\mathcal{X}}(\Gamma_i, V_i)$ is the actual space of states determined from the measurement and, therefore, corresponds to $\mathcal{H}^{\rm mes}(\Gamma_i)$ in the measurement-based description (\ref{ortps}) where the state
$\ket{\psi_i}_{\Gamma_i} \in \mathcal{H}_{\mathcal{X}}(\Gamma_i, V_i)$ is identified with 
$\ket{\psi_i}_{\Gamma_i} \in \mathcal{H}^{\rm mes}(\Gamma_i)$.

\begin{figure}[t]
 \vspace*{-5mm}\hspace*{0mm}
 \begin{center}
  \includegraphics[width=2.7in]{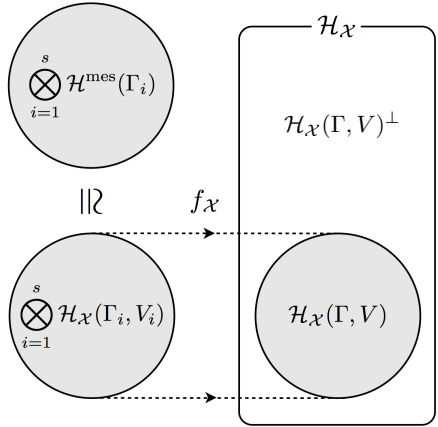}
 \end{center}
 \vspace*{-2mm}
 \caption{A diagrammatical representation of the spaces mentioned in (\ref{mapsp}) and (\ref{spodec}).  In the total space $\mathcal{H}_{\mathcal{X}}$, we find the subspace $\mathcal{H}_{\mathcal{X}}(\Gamma, V)$ isomorphic to $\otimes_{i} \mathcal{H}_{\mathcal{X}}(\Gamma_i, V_i)$ which has a TPS.  The latter is then identified with the space $\mathcal{H}^{\rm mes}$ describing the measurement outcomes.}
 \label{f1}
\end{figure}

From the description for the subsystems, we obtain the Hilbert space of the total system by
\begin{equation}
\mathcal{H}_{\mathcal{X}}(\Gamma, V)
= \left[\bigotimes_{i=1}^s \mathcal{H}_{\mathcal{X}}(\Gamma_i, V_i) \right]_{\mathcal{X}}.
  \label{totsppro}
\end{equation}
Note that, due to the  (anti)symmetrization $\mathcal{X}$, the resultant space $\mathcal{H}_{\mathcal{X}}(\Gamma, V)$ has no TPS with respect to the decomposition $\Gamma$ and, hence, no obvious correspondence with $\mathcal{H}^{\rm mes}$ in (\ref{ortps}).
In spite of this, the two spaces can be made
isomorphic based on the identification $\mathcal{H}_{\mathcal{X}}(\Gamma_i, V_i) \cong \mathcal{H}^{\rm mes}(\Gamma_i)$ mentioned above. 
Indeed, it is attained, with this identification, by the map,
\be
f_{\mathcal{X}} : \,  \mathcal{H}^{\rm mes} \cong \bigotimes_{i=1}^s \mathcal{H}_{\mathcal{X}}(\Gamma_i, V_i) 
\mapsto \mathcal{H}_{\mathcal{X}}(\Gamma, V),
  \label{mapsp}
\ee
defined by
\be
f_{\mathcal{X}}\left( \bigotimes_{i=1}^s\ket{\psi_i}_{\Gamma_i} \right)  = \sqrt{M}\mathcal{X}\bigotimes_{i=1}^s\ket{\psi_i}_{\Gamma_i},
\label{corsp}
\ee
with the normalization factor $M:=N!/\prod_{i}^s |\Gamma_i|!$.   Obviously, the map $f_{\mathcal{X}}$ is surjective by construction, and to see  
the injectivity, we note that, thanks to the orthogonal structure $V$ in (\ref{ortst}), the innerproduct is invariant
under the map \cite{footnote1}.  It follows that  
$\| \otimes_{i}\ket{\psi_i}_{\Gamma_i}\| = \| \sqrt{M}\mathcal{X}\otimes_{i}\ket{\psi_i}_{\Gamma_i} \|$,
which ensures the injectivity of the map and hence the isomorphism  (see Fig.{\ref{f1}}).   

The isomorphism (\ref{mapsp}) induces the correspondence between the observables in the two spaces.  If $O_i$ are the observables
in $\mathcal{H}_{\mathcal{X}}(\Gamma_i, V_i)$ for $i = 1, \ldots, s$, then the observable in $\mathcal{H}_{\mathcal{X}}(\Gamma, V)$ for simultaneous measurement reads
\be
\widetilde O =  M \mathcal{X} \(\bigotimes_{i=1}^{s} O_i \) \mathcal{X}.
\label{orprod}
\ee
The set of all such operators defines a subset ${\cal T}(\Gamma, V)$ of observables in $\mathcal{H}_{\mathcal{X}}(\Gamma, V)$.  
With the identification of the observables $O_i$ between $\mathcal{H}_{\mathcal{X}}(\Gamma_i, V_i)$ and 
$\mathcal{H}^{\rm mes}(\Gamma_i)$,  
the induced isomorphism for the observables corresponding to (\ref{corsp}) is given (by abusing the symbol) by
\be
f_{\mathcal{X}}\left(\bigotimes_{i=1}^{s} O_i \right)  = M \mathcal{X} \(\bigotimes_{i=1}^{s} O_i \) \mathcal{X}.
\ee
This also implies the isomorphism between ${\cal T}^{\rm mes}$ and ${\cal T}(\Gamma, V)$ through the correspondence
$\widehat O \leftrightarrow \widetilde O$.

In the provision-based description, the criterion on the entanglement of indistinguishable particles then emerges as follows.  
Given an arbitrary (normalized) state $\ket{\Psi} \in \mathcal{H}_{\mathcal{X}}$, we first decompose it as 
\be
\ket{\Psi} = \ket{\Psi(\Gamma,V)} + \ket{\Psi(\Gamma,V)^\perp},
\label{odecst}
\ee
according to the orthogonal decomposition,
\be
\mathcal{H}_{\mathcal{X}} 
=\mathcal{H}_{\mathcal{X}}(\Gamma, V) \oplus\ \mathcal{H}_{\mathcal{X}}(\Gamma, V)^{\perp}.
\label{spodec}
\ee
Since $\ket{\Psi(\Gamma,V)^\perp}$ has a vanishing support for the observables in ${\cal T}(\Gamma, V)$ and is filtered out by the measurement, the only part significant for correlation is $\ket{\Psi(\Gamma,V)}$.
Thus, for studying correlations in the measurement outcomes ignoring the events which are not detected in the setup, 
one uses the remormalized state $\| \ket{\Psi(\Gamma,V)}\|=1$.  
We now see that, if the observable part takes the form,
\begin{equation}
\ket{\Psi(\Gamma,V)}  = 
 \sqrt{M}\mathcal{X}\bigotimes_{i=1}^s\ket{\psi_i}_{\Gamma_i},
 \label{pfacto}
\end{equation}
then for any $\widetilde O \in {\cal T}(\Gamma, V)$ we have the factorization,
\begin{eqnarray}
 \bra{\Psi}  \widetilde{O} \ket{\Psi} = \prod_{i=1}^s\bra{\Psi}    \widetilde O_i\ket{\Psi},
  \label{efacto}
\end{eqnarray}
in analogy with (\ref{facto}).  Since the converse is also true, we learn that the state $\ket{\Psi}$ is separable if and only if the part
$\ket{\Psi(\Gamma,V)}$ in (\ref{odecst}) admits the (anti)symmetrized direct product form (\ref{pfacto}); if not, it is entangled.
In more simple terms, to examine the separability of a given state 
$\ket{\Psi}  \in \mathcal{H}_{\mathcal{X}}$, we just concentrate on the observable part $\ket{\Psi(\Gamma,V)}$ and then strip it off the projection $\mathcal{X}$ (and perform necessary renormalization) to obtain, via the identification in Fig.{\ref{f1}}, the corresponding state $\ket{{\Psi}^{\rm mes}(\Gamma,V)} \in \mathcal{H}^{\text{mes}}$ describing the measurement outcomes directly.    In the case (\ref{pfacto}) we find $\ket{{\Psi}^{\rm mes}(\Gamma,V)} = \otimes_{i=1}^s\ket{\psi_i}_{\Gamma_i}$, which is factorizable and hence separable.
As is evident from the explicit dependence on $V$,  the entanglement of the state is determined only relatively with respect to the measurement setup.   

Generalization of our argument to mixed states is straightforward.   Given a density matrix $\rho$ on ${\cal H}_{\cal X}$,
one can decompose it as
\be
\rho=\begin{pmatrix}
\rho(\Gamma, V)      &   * \\
*      &  \rho(\Gamma, V)^\perp
\end{pmatrix},
\ee
where $\rho(\Gamma, V)$ and $\rho(\Gamma, V)^\perp$ are (unnormalized) density matrices on ${\cal H}_{\cal X}(\Gamma, V)$ and ${\cal H}_{\cal X}(\Gamma, V)^\perp$, respectively.  By virture of the isomorphism $f_{\cal X}$, the separability criterion for the mixed distinguishable systems \cite{Werner} can be utilized for 
the density matrix $\rho^{\rm mes}(\Gamma,V)$ which is defined from $f_{\cal X}^{-1}\(\rho(\Gamma,V)\)$ with a suitable rescaling to fulfill $\Tr\, \rho^{\rm mes}(\Gamma,V) = 1$. 
We then find that a mixed state $\rho$ on ${\cal H}_{\cal X}$ is separable under our measurement setup specified by $\Gamma$ and $V$ if $\rho^{\rm mes}(\Gamma,V)$ admits the form, 
\be
\rho^{\rm mes}(\Gamma,V)=\sum_\alpha p_\alpha\ket{\Psi^{\rm mes}_\alpha}\bra{\Psi^{\rm mes}_\alpha},
\ee
where $\ket{\Psi_\alpha^{\rm mes}}\in{\cal H}^{\rm mes}$ are separable pure states and $\{p_\alpha\}$ satisfies $\sum_\alpha p_\alpha=1$ and $p_\alpha\ge0$.

It should be clear by now that since a state of indistinguishable particles, either it is pure or mixed, can be mapped to a state in ${\cal H}^{\rm mes}$, the entanglement of the state can be studied in terms of the standard entanglement measures developed for distinguishable particles.   This will be demonstrated next.

\section{Measurement setup dependence of entanglement: examples}

\begin{figure}[t]
 \includegraphics[width=5cm,angle=270]{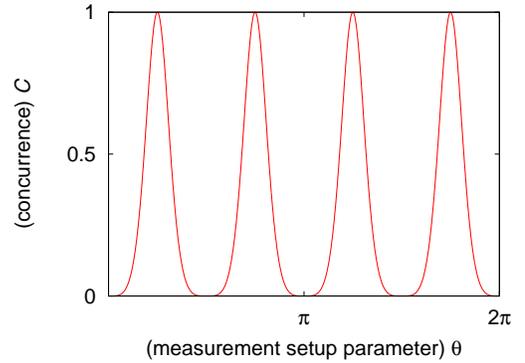}
 \caption{The concurrence $C$ of $\ket{{\Psi}^{\text{mes}}(\Gamma,V)}$ as a function of the angle $\theta$.  The variation shows that the state undergoes the change from complete separation to maxmial entanglement by the adjustment of the setup angle $\theta$.
}
 \label{concu1}
\end{figure}

In order to evaluate explicitly the dependence of entanglement on the measurement setup we choose, let us 
consider the case of $N=2$ fermions, each possessing $n=4$ dimensional constituent space, given in the state,
\begin{equation}
 \ket{\Psi} = \sqrt{2}\mathcal{A}\ket{e_1}_1\ket{e_3}_2 \in \mathcal{H}_{\mathcal{A}}.
 \label{ferexample}
\end{equation}
We wish to examine if,  and to what extent, the state is entangled under the 
partition $\Gamma=\{\{1\},\{2\}\}$ when our setup is \lq rotated\rq\ among the set of basis
$\{\ket{e_1},\ket{e_4}\}$ and $\{\ket{e_2},\ket{e_3}\}$.  To this end, we 
adopt the orthogonal decomposition (which defines the measurement setup)
$V= \{V_1, V_2\}$ with
\begin{equation}
 V_1 = \text{span}\{\ket{e'_1},\ket{e'_2}\},
 \quad
 V_2 = \text{span}\{\ket{e'_3},\ket{e'_4}\},
\end{equation}
where
\begin{eqnarray}
&&\!\!\!\!\!\!   \ket{e'_1} = c\ket{e_1} + s\ket{e_4},  \quad \ket{e'_2} =-s\ket{e_3} + c\ket{e_2},
 \nonumber \\
&&\!\!\!\!\!\!  \ket{e'_3} = c\ket{e_3} + s\ket{e_2}, \quad \ket{e'_4} = -s\ket{e_1} + c\ket{e_4},  
\end{eqnarray}
and we have used the shorthand $c=\cos\theta$, $s=\sin\theta$ to express the rotation with angle $\theta$.  
According to the decomposition (\ref{odecst}),  the measurable part turns out to be
\begin{equation}
\ket{\Psi(\Gamma,V)}
	= \mathcal{A} \(c^2\ket{e'_1}_1\ket{e'_3}_2 - s^2\ket{e'_2}_1\ket{e'_4}_2\),
\end{equation}
up to a constant.
We then map it to the corresponding normalized state in $\mathcal{H}^{\text{mes}}$ as
\begin{equation}
\label{example-mes-space}
 \ket{{\Psi}^{\text{mes}}(\Gamma,V)}
	= \frac{1}{\sqrt{c^4+s^4}} \(c^2\ket{e'_1}_1\ket{e'_3}_2 - s^2\ket{e'_2}_1\ket{e'_4}_2\).
\end{equation}
The amount of entanglement may be evaluated by the (squared) concurrence,
\begin{eqnarray}
 C\( \ket{{\Psi}^{\rm mes}(\Gamma,V)}\)  
&=& 2\[1-\Tr_1 \(\Tr_2 \ket{{\Psi}^{\text{mes}}} \bra{{\Psi}^{\text{mes}}} \)^2 \]
 \nonumber \\
&=& \frac{4}{(\tan^2\theta+\cot^2\theta)^2},
\end{eqnarray}
and the result is depicted in Fig. \ref{concu1}.   We find that the state (\ref{ferexample}) is strictly separable at 
$\theta = n\pi/2$ and maximally entangled at $\theta = (n + 1/2)\pi/2$ for integer $n$, and it can take any intermediate values $C$ in between.

\begin{figure}[t]
\begin{center}
  \includegraphics[width=7cm]{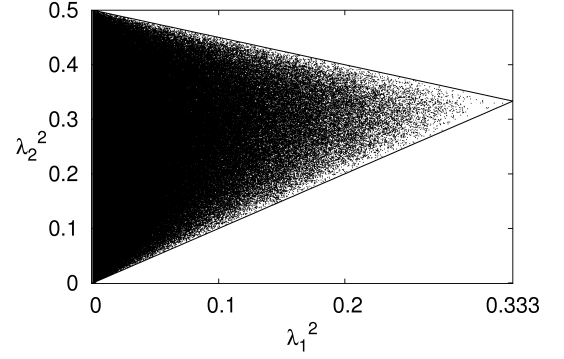}
 \caption{The distribution of the  Schmidt coefficients of the state (\ref{bosstmes}) with the ordering $\lambda_1\leq \lambda_2\leq \lambda_3$ ($\lambda_3$ is not shown because it can be determined from other two) obtained under random setups of measurement  for the state (\ref{bosexample}) provided by $10^6$ distinct unitaries $U$.   The diagonal lines represent $\lambda_1=\lambda_2$ and $\lambda_2=\lambda_3$, whose intersection corresponds to the maximally entangled state.   The points fill out basically the entire region of the triangle, although in our random generation the density becomes scarcer for states which are highly entangled.}
 \label{n6V3V3666666}
\end{center}
\end{figure}

As a second example, we consider the case of $N=2$ bosons with $n= 6$ prepared in the state, 
\begin{equation}
 \ket{\Psi}=\frac{1}{\sqrt{6}}\sum_{i=1}^6 \ket{e_i}_1\ket{e_i}_2 \in \mathcal{H}_{\mathcal{S}}.
  \label{bosexample}
\end{equation}
As before, we study the entanglement for the 
partition  $\Gamma=\{\{1\},\{2\}\}$ when the \lq rotated\rq\ family of entanglement setups are considered, which are now provided by   
\begin{equation}
 V_1 = \text{span}\{\ket{e'_1},\ket{e'_2}, \ket{e'_3}\}, \quad
 V_2 = \text{span}\{\ket{e'_4},\ket{e'_5},\ket{e'_6}\},
\end{equation}
with 
\begin{equation}
\ket{e'_i}=U\ket{e_i}, \quad U\in {\rm U}(6).
\label{ubasisbos}
\end{equation}
Analogously to Eq. (\ref{example-mes-space}), we can find the corresponding state $\ket{{\Psi}^{\text{mes}}(\Gamma,V)}$ in $\mathcal{H}^{\text{mes}}$.
This time, however, instead of simply evaluating the concurrence we study the extent of variation in the state $\ket{{\Psi}^{\text{mes}}(\Gamma,V)}$ that can arise 
under distinct measurement setups obtained by altering the unitary matrix
$U$ in (\ref{ubasisbos}).   To do this, we first implement the Schmidt decomposition for the state $\ket{{\Psi}^{\text{mes}}(\Gamma,V)}$ as 
\begin{equation}
\ket{{\Psi}^{\text{mes}}(\Gamma,V)} = \sum_{i=1}^3 \lambda_i \ket{e''_i}_1\ket{e''_{i+3}}_3, \quad 
\sum_{i=1}^3 \lambda_i^2=1,
  \label{bosstmes}
\end{equation}
where $\{\ket{e''_{i}}\}_{i=1}^3$, $\{\ket{e''_{i}}\}_{i=4}^6$ are the Schmidt bases $\{\ket{e''_{i}}\}_{i=1}^3$, $\{\ket{e''_{i}}\}_{i=4}^6$ each defined within the measurable subspaces $V_1$, $V_2$.    We then observe the distribution of states which is invariant under local unitary operations from the distribution of Schmidt coefficients.
Fig.\ref{n6V3V3666666} shows the values of $\lambda_i$ for $i = 1, 2$ obtained by a random generation of $U$,  which suggests
that by tuning $U$ properly the state (\ref{bosexample}) can furnish virtually any possible states which are discriminable by the Schmidt coefficients.  

Despite the relative nature of entanglement with respect to the measurement setup, there exists a special class of states in the bosonic case
which are separable under all measurement setups.   These are the i.i.d.~pure states $ \ket{\Psi}  \in \mathcal{H}_{\mathcal{S}}$ defined by 
\begin{equation}
 \ket{\Psi} = \bigotimes_{k=1}^N\ket{\phi}_k, \qquad \ket{\phi}_k \in{\cal H}_k.
 \label{iidst}
\end{equation}
To see the universal separability of the state, we decompose $\ket{\phi}_k$ 
according to Eqs.(\ref{ortst}) and (\ref{addone}) as
\be
 \ket{\phi}_k = \sum_{i=0}^s \ket{\varphi_i}_k,
 \qquad
 \ket{\varphi_i}_k \in V_i(\mathcal{H}_k).
 \ee
Plugging this into Eq.(\ref{iidst}), we obtain Eq.(\ref{odecst}) with
\be
\!\!\! \ket{\Psi(\Gamma,V)} = \sqrt{M}\mathcal{S}\bigotimes_{i=1}^s \ket{\psi_i}_{\Gamma_i},
\ee
where
\be
\ket{\psi_i}_{\Gamma_i} \propto \bigotimes_{k\in \Gamma_i}\ket{\varphi_i}_k.
\ee
Since the part $\ket{\Psi(\Gamma,V)}$, if non-vanishing, belongs to the class Eq.(\ref{pfacto}), we see that the i.i.d.~states (\ref{iidst}) are separable.  Further, since this is true for any choice of $(\Gamma,V)$, the separability holds irrespective of the measurement setup.   Interestingly, one finds that, for $N=2$, $n\geq 4$, the converse is also true: states which are universally separable must be the i.i.d.~states.

\section{Conclusion and Discussions}

In summary, we have presented a general criterion for entanglement of an indistinguishable $N$ particle system 
decomposed into $s$ subsystems based on the unambiguous measurability of correlation.  The point is that, although 
the Hilbert space $\mathcal{H}_{\mathcal{X}}$ of the system does not admit a TPS, one can find 
a subspace $\mathcal{H}_{\mathcal{X}}(\Gamma, V) \subset \mathcal{H}_{\mathcal{X}}$ 
which has a TPS and is directly related to the space $\mathcal{H}^{\rm mes}$ describing the measurement outcomes.  
Since $\mathcal{H}^{\rm mes}$ has a common structure with the space of distinguishable particles, our approach allows us to treat indistinguishable particles on the equal basis with distinguishable ones. Consequently, the handling of states without considering the effect of (anti)symmetrization practiced regularly in quantum optics is found to be safe as long as it deals with the space $\mathcal{H}^{\rm mes}$. 

 The structure of ${\cal H}^{\rm mes}$ also implies that the standard measures of entanglement devised so far can be used equally for the indistinguishable case;  for instance, the monotonicity of entanglement measures with respect to local operations and classical communications (LOCC) is preserved under the mapping $f_{\cal X}$.  
 This is shown by observing that
all ingredients of LOCC for distinguishable particles \cite{Vidal00} have their counterparts in ${\cal H}_{\cal X}(\Gamma, V)$ provided by the application of $f_{\cal X}$. Since generalized measurements, POVM, can be built from some of the ingredients of LOCC (Naimark\rq s theorem \cite{Peres93}), the mapping $f_{\cal X}$ induces the analogues of generalized local measurements in ${\cal H}_{\cal X}(\Gamma, V)$.

As stated in the introduction, for bosonic systems the characterization of separability has not been done uniquely in the literature, and in fact it is mentioned in \cite{GM04}  (Theorem 3.6) that there are states which are separable from the criterion of \cite{GMW02, GM03} but entangled from that of \cite{SCKLL01}.    We observe, however, that 
the theorem used is based on a unitary transformation between two different bases of the one-particle state space, which amounts to a change in the orthogonal structure $V$ in our language.  As explicitly shown in the previous section, such a change gives rise to states with different amounts of entanglement observed in the altered settings, and in this sense one can regard the apparent discrepancy as just a reflection of the relative nature of entanglement.  

{}Finally, we mention that our approach can also be applied, rather trivially,  to a system consisting of both bosons and fermions.   Indeed, we may first treat bosons and fermions separately in the provisional spaces ${\cal H}_{\cal S}$ and ${\cal H}_{\cal A}$ in our approach,  and then combine them together to form the total provisional space ${\cal H}_{\cal S}\ot{\cal H}_{\cal A}$.   Since ${\cal H}_{\cal S}$ and ${\cal H}_{\cal A}$ have their own isomorphism $f_{\cal S}$ and $f_{\cal A}$ defined from the orthogonal structures equipped with them, it is evident that through the combination of $f_{\cal S}^{-1}$ and $f^{-1}_{\cal A}$ we obtain the corresponding state space which has a TPS and can be identified with the total measurement space ${\cal H}^{\rm mes}$.   This reasoning can be extended in principle to more complex systems consisting of several distinct species of fermions and bosons, {\it e.g.}, those describing interactions between matter and gauge mediators such 
 as photons and gluons.

\begin{acknowledgments}
T.S. is supported by JSPS Research Fellowships for Young Scientists, T.I. is supported by \lq Open Research Center\rq~Project for Private Universities, and I.T. is supported by the Grant-in-Aid for Scientific Research (C), No.~20540391-H22, all of MEXT, Japan. 
\end{acknowledgments}



\begin{thebibliography}{99}
%
\bibitem{EPR}
{A. Einstein, B. Podolsky and N. Rosen, Phys. Rev.
{\bf 47}, 777 (1935).}
%
\bibitem{Sch}
E. Schr{\" o}dinger, 
{\it Proceedings of the Cambridge Philosophical Society} {\bf 31}, 555 (1935); {\bf 32}, 446 (1936).
%
\bibitem{WC63}
E. H. Wichmann and J. H. Crichton,
Phys. Rev. {\bf 32}, 2788 (1963);
%
S. Weinberg,
{\em Quantum Theory of Fields, Volume I, Foudations}, 
(Cambridge university press, Cambridge, 1995).
%
\bibitem{Peres93}
A. Peres,
{\em Quantum Theory: Concepts and Methods}, 
(Kluwer, Dordrecht, 1993).
%
\bibitem{GMW02}
G.-C. Ghirardi, L. Marinatto and T. Weber, 
{J. Stat. Phys.} {\bf 108}, 49 (2002).
%
\bibitem{GM03}
G.-C. Ghirardi and L. Marinatto, 
{Fortschr. Phys.} {\bf 51}, 379 (2003).
%
\bibitem{Zanardi01}
P. Zanardi, 
Phys. Rev. Lett. {\bf 87}, 077901 (2001);
%
P. Zanardi, D. A. Lidar and S. Lloyd,
{\it ibid.} {\bf 92}, 060402 (2004).
%
\bibitem{SCKLL01}
J. Schliemann {\it et al.},
Phys. Rev. A. {\bf 64}, 022307 (2001);
%
K. Eckert {\it et al.},
Ann. Phys. {\bf 299}, 88 (2002).
%
\bibitem{PY01}
P. Pa{\v s}kauskas, and L. You,
Phys. Rev. A. {\bf 64}, 042310 (2001);
%
Y.-S. Li {\it et al.},
{\it ibid.} {\bf 64}, 054302 (2001);
%
H. M. Wiseman, and John A. Vaccaro,
Phys. Rev. Lett. {\bf 91}, 097902 (2003).
%
\bibitem{GM04}
G.-C. Ghirardi and L. Marinatto, 
Phys. Rev. A. {\bf 70}, 012109 (2004).
%
\bibitem{IST10}
{T. Ichikawa, T. Sasaki and I. Tsutsui, 
J. Math. Phys. {\bf 51}, 062202 (2010).}
%
\bibitem{footnote1}
{The proof is given in Ref.\cite{IST10} for the fermionic case, but the bosonic case can also be dealt with analogously.
}
%
\bibitem{Werner}
{R. F. Werner, 
Phys. Rev. A. {\bf40}, 4277 (1989).}
%
%
\bibitem{Vidal00}
{G. Vidal, 
J. Mod. Opt. {\bf47}, 355 (2000).}
\end{thebibliography}
\end{document}